\documentclass[aps,prd,reprint,twocolumn,groupedaddress,superscriptaddress,floatfix]{revtex4-2}
\usepackage{lipsum}
\usepackage{graphicx} 
\usepackage{dcolumn}  
\usepackage{bm}       
\usepackage{amssymb}   
\usepackage{amsmath}
\usepackage{xcolor}
\usepackage{amsfonts}
\usepackage{hyperref}
\hypersetup{colorlinks=true, linkcolor=blue, citecolor=blue, urlcolor=blue}
\usepackage{bm}
\usepackage{xcolor}
\begin{document}
\title{Data-Driven Einstein-Dilaton Model for Pure Yang-Mills Thermodynamics and Glueball Spectrum}

\author{Xun Chen}
\email{chenxun@usc.edu.cn}
\affiliation{School of Nuclear Science and Technology, University of South China, Hengyang 421001, China}
\affiliation{INFN --- Istituto Nazionale di Fisica Nucleare --- Sezione di Bari Via Orabona 4, 70125, Bari, Italy}

\author{Yidian Chen}
\email{chenyidian@hznu.edu.cn}
\affiliation{School of Physics, Hangzhou Normal University, Hangzhou, 311121, China}

\author{Kai Zhou}
\email{zhoukai@cuhk.edu.cn}
\affiliation{School of Science and Engineering, The Chinese University of Hong Kong, Shenzhen (CUHK-Shenzhen), Guangdong, 518172, China}
\affiliation{School of Artificial Intelligence, Chinese University of Hong Kong (Shenzhen), Shenzhen, Guangdong, China}

\date{\today}

\begin{abstract}
\noindent We develop a machine learning assisted holographic model that consistently describes both the equation of state and glueball spectrum of pure Yang-Mills theory, achieved through neural network reconstruction of Einstein-dilaton gravity. Our framework incorporates key non-perturbative constraints of lattice QCD data: the ground ($0^{++}$) and first-excited ($0^{++*}$) scalar glueball masses pins down the infrared (IR) geometry, while entropy density data anchors the ultraviolet (UV) behavior of the metric. A multi-stage neural network optimization then yields the full gravitational dual—warp factor $A(z)$ and dilaton field $\Phi(z)$—that satisfies both spectroscopic and thermodynamic constraints. The resulting model accurately reproduces the deconfinement phase transition thermodynamics (pressure, energy density, trace anomaly) and predicts higher glueball excitations ($0^{++**}$, $0^{++***}$) consistent with available lattice calculations. This work establishes a new paradigm for data-driven holographic reconstruction, solving the long-standing challenge of unified description of confinement thermodynamics and spectroscopy. 
\end{abstract}

\maketitle
\section{Introduction} 
Quantum Chromodynamics (QCD) provides the fundamental description of strong interactions, yet its non-perturbative regime remains notoriously challenging to explore. In the pure gluon sector (SU(3) Yang-Mills theory), confinement reveals two hallmark features: a markedly non-ideal equation of state (EoS) near the deconfinement transition and the emergence of glueballs—exotic states composed solely of gauge fields. Lattice calculations have provided crucial insights into the thermodynamics \cite{Boyd:1996bx,HotQCD:2014kol}, scattering cross section \cite{Yamanaka:2019aeq,Yamanaka:2019yek}, and scalar glueball spectrum \cite{deGier:2016gcy,Athenodorou:2020ani,Chen:2005mg}; however, these methods face computational limitations in studying real-time dynamics, finite-density regimes, and higher excited states.

Holographic approaches based on the AdS/CFT correspondence \cite{Maldacena:1997re} offer a complementary, non-perturbative framework for investigating strongly-coupled gauge theories. Pioneering comparisons between holographic QCD models and lattice thermodynamics were carried out in the early works \cite{Gursoy:2008bu, Gubser:2008yx}. Einstein-dilaton gravity models have successfully described the EoS near the deconfinement transition \cite{Gursoy:2007cb,Li:2011hp,He:2013qq,Yang:2014bqa,Toniato:2025gts,Dudal:2021jav,Bohra:2019ebj,Rougemont:2023gfz,Grefa:2021qvt,He:2022amv,Zhao:2023gur,Zhao:2022uxc}, while bottom-up holographic constructions \cite{Karch:2006pv,Colangelo:2007pt,Bellantuono:2015fia,FolcoCapossoli:2015jnm,FolcoCapossoli:2016fzj,FolcoCapossoli:2016uns,Rodrigues:2016cdb,Rodrigues:2016kez, FolcoCapossoli:2019imm,Capossoli:2021ope,Zhang:2022uin,Chen:2022goa,Zhao:2021ogc,Ballon-Bayona:2017sxa,Ballon-Bayona:2021tzw,Chen:2025ncu} capture the scalar glueball spectrum. However, existing models typically focus exclusively on either thermodynamics or spectroscopy, no unified single holographic model has yet been constructed to satisfy both lattice benchmarks simultaneously. The obstacle is the inverse problem: reconstructing a self-consistent bulk geometry and dilaton potential directly from QCD data is analytically intractable.

To bridge this gap, we develop a Neural Network Holographic Model (NNHM) that integrates both scalar-glueball spectrum and thermodynamic observables within a single Einstein-dilaton background. In recent years, machine learning has also emerged as a powerful tool widely used in high-energy physics \cite{Huang:2025uvc,Li:2025csc,Pang:2024kid,Zhou:2023pti,Li:2022ozl,Aarts:2025gyp,Li:2025obt,Soma:2022vbb,Zhang:2025juc,Zhang:2022uqk,Zhou:2018ill,Jiang:2021gsw,Wang:2023exq}. Our approach builds upon recent advances in physics-informed machine learning methods, which have already proven effective in reconstructing consistent gravitational duals \cite{Hashimoto:2018ftp, Hashimoto:2018bnb, Akutagawa:2020yeo, Hashimoto:2021ihd, Hashimoto:2024yev,Song:2020agw,Ahn:2024gjf,Ahn:2024jkk,Ahn:2025tjp,Fu:2024wkn,Cai:2024eqa,Chang:2024ksq,Chen:2024ckb,Chen:2024mmd,Mansouri:2024uwc,Luo:2024iwf,Dai:2025dir}, to solve the inverse problem in Einstein-dilaton gravity. We train parallel neural networks to incorporate lattice data for the ground state ($0^{++}$) and first excited state ($0^{++*}$) scalar glueball masses which constrains the IR ($z \to \infty$) geometry, and simultaneously use entropy density data to calibrate the UV ($z \to 0$) behavior of the metric. Through a systematically designed multi-stage optimization in merging both the two constraints, we reconstruct the full self-consistent warp factor $A(z)$ and dilaton field $\Phi(z)$, achieving the first holographic dual that consistently describes both the confinement thermodynamics and glueball spectrum.
\section{Review of EMD framework} 
\begin{figure}
    \centering
    \includegraphics[width=8cm]{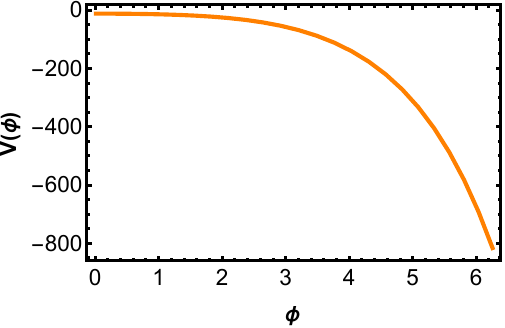}
    \caption{The scalar potential $V(\phi)$ as a function of the scalar field $\phi$ in our model at vanishing temperature and chemical potential. } 
    \label{vphi}
\end{figure}

Firstly, we review the 5-dimensional Einstein-Dilaton systems at finite temperature \cite{Li:2011hp,He:2013qq,Yang:2014bqa,Yang:2015aia,Dudal:2017max,Dudal:2018ztm,Chen:2018vty,Chen:2020ath,Zhou:2020ssi,Chen:2019rez,Chen:2024ckb,Chen:2024epd,Chen:2024mmd,Zhu:2025gxo}. In this work, our construction follows the “bottom-up” strategy in which one specifies a physically motivated warp–factor profile and then reconstructs the corresponding dilaton potential. This is conceptually different from the Refs. \cite{Gursoy:2007er,Gursoy:2007cb,Gubser:2008ny}, where the dilaton potential is postulated from the beginning and the geometry is obtained as a solution of the Einstein–dilaton equations. The action is expressed by the following equation:
\begin{equation}
\begin{aligned}
S_{E} =\frac{1}{16 \pi G_5} \int \mathrm{d}^5 x \sqrt{-g}\left[R-\frac{1}{2} \partial_\mu \Phi \partial^\mu \Phi-V(\Phi)\right].
\end{aligned}
\end{equation}
$V(\Phi)$ is the dilaton potential, and $G_5$ is the Newton constant in five dimensions.
The ansatz of metric in the Einstein frame is
\begin{equation}
d s^2=\frac{L^2 e^{2 A(z)}}{z^2}\left[-g(z) d t^2+\frac{d z^2}{g(z)}+d \vec{x}^2\right],
\end{equation}
where $z$ is the holographic coordinate in the fifth dimension, and the AdS radius $L$ is set to one, i.e., $L = 1 \, \mathrm{GeV}^{-1}$.
The boundary conditions are specified as follows: at the horizon $z = z_h$,
\begin{equation}
g(z_h) = 0.
\end{equation}
As we approach the boundary ($z \rightarrow 0$), we impose that the metric in the string frame asymptotically approaches $\mathrm{AdS}_5$, corresponding to the boundary conditions
\begin{equation}
A(0) = \Phi(0) = 0, \quad g(0) = 1.
\end{equation}
$g(z)$, $\phi(z)$ and $V(z)$ have following form
\begin{equation}
\begin{aligned}
g(z) &= 1 - \frac{\int_0^z y^3 e^{-3 A} \, \mathrm{d} y }{\int_0^{z_h} y^3 e^{-3 A} \, \mathrm{d} y}, \\
\Phi^{\prime}(z) &= \sqrt{6 \left( A^{\prime 2} - A^{\prime \prime} - \frac{2 A^{\prime}}{z} \right)}, \\
V(z) &= -\frac{3 z^2 g e^{-2 A}}{L^2} \bigg[ A^{\prime \prime} + A^{\prime} \left( 3 A^{\prime} - \frac{6}{z} + \frac{3 g^{\prime}}{2 g} \right) \\
&\quad - \frac{1}{z} \left( -\frac{4}{z} + \frac{3 g^{\prime}}{2 g} \right) + \frac{g^{\prime \prime}}{6 g} \bigg].
\end{aligned}
\end{equation}
The boundary condition $\Phi(0)=0$ fixes the scalar source in the dual theory for $\Delta = 4$. This eliminates the scalar charge term in the first law. In this work, $A(z)$ is represented by a neural network, whose form is constrained by the data. The potential of the scalar $V(\phi)$ in our model is numerically solved in Fig. \ref{vphi}.
The Hawking temperature and entropy of the black hole solution are given by
\begin{equation}
T = \frac{\kappa}{2\pi}= \frac{1}{4\pi} \left|  \frac{dg}{dz} \right|_{z = z_h} =\frac{z_h^3 e^{-3 A(z_h)}}{4 \pi \int_0^{z_h} \mathrm{d} y  y^3 e^{-3 A(y)}},
\end{equation}
where $\kappa$ is the surface gravity. The entropy density is defined as 
\begin{equation}
s = \frac{S}{V_3} = \frac{A_h}{4 G_5 V_3} = \frac{e^{3 A(z_h)}}{4 G_5 z_h^3}.
\end{equation}
In this expression \cite{Natsuume:2014sfa}, $A_h$ is the area of the black hole horizon, $V_3$ is the spatial three-volume, and $G_5$ denotes the five-dimensional Newton constant, which will be determined later when fitting the lattice QCD. The free energy can then be calculated as
\begin{equation}
F = -\int s  \mathrm{d}T,
\end{equation}
where the integration is performed over the temperature. The pressure is defined as $p = -F$. Finally, the energy density of the system is derived from the thermodynamic relation:
\begin{equation}
\epsilon = -p + s T.
\end{equation}
A holographic model is dimensionless. To introduce a physical unit, we need to match model predictions with experimental data or lattice QCD results. In our work, we fix the scale by matching the critical temperature to its physical value in GeV. Once this scale is set, we compute the hadron mass spectrum in the same physical unit (GeV), obtaining results consistent with lattice QCD.

\section{Reconstruction of glueball potential}
\begin{figure}
    \centering
    \includegraphics[width=9cm]{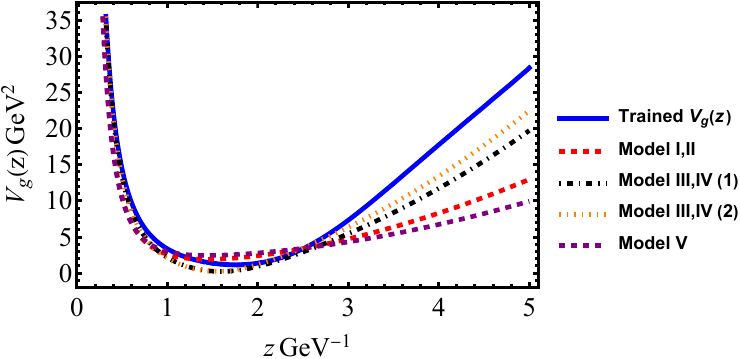}
    \caption{ Comparison between the extracted glueball potential $V_g(z)$ from lattice QCD data and predictions from other holographic QCD models \cite{Zhang:2021itx}. }
    \label{potential_plot}
\end{figure}

\begin{figure}
    \centering
    \includegraphics[width=8cm]{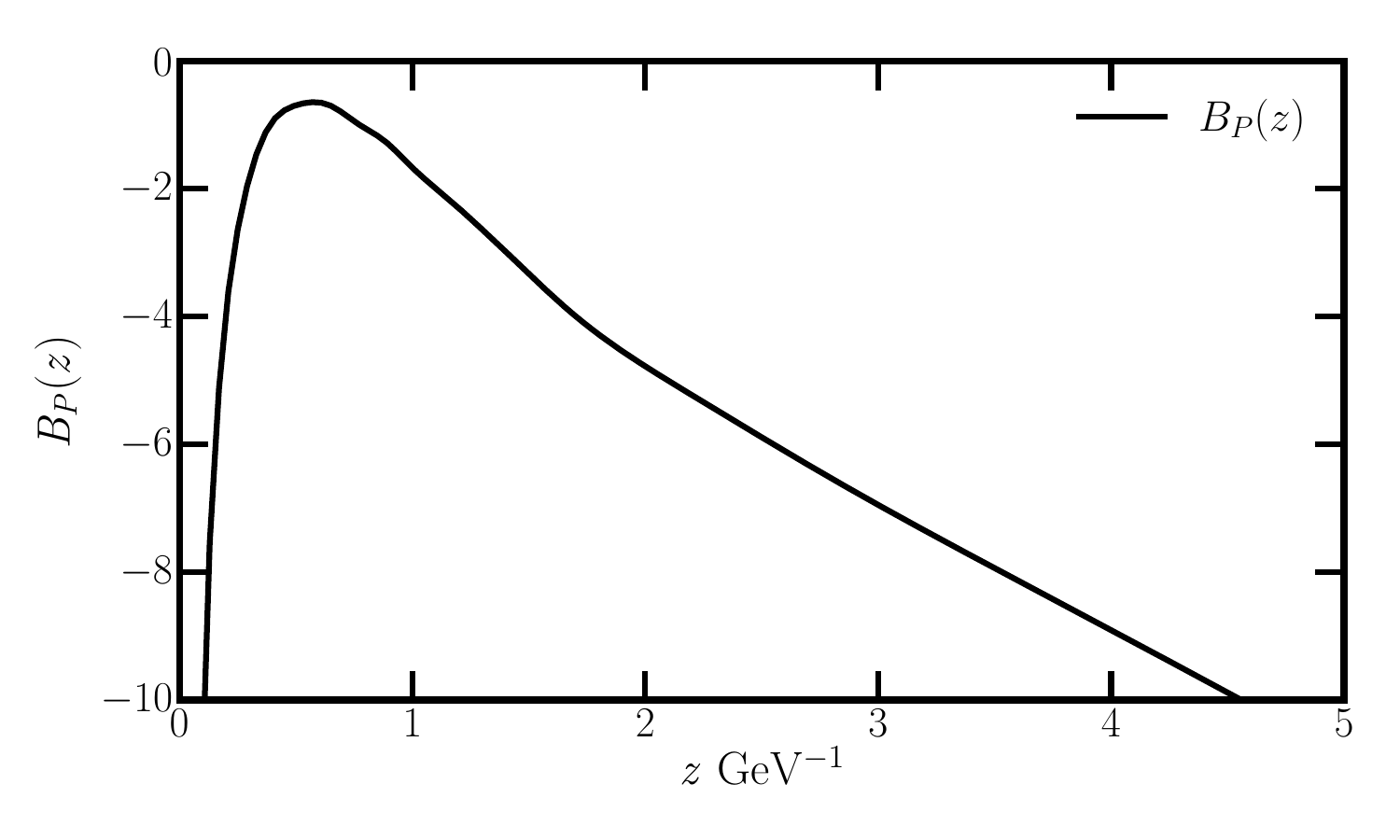}
    \caption{Reconstructed composite derivative $B_P(z) = 3A_s'(z) - \Phi'(z)$ obtained by solving Eq. \eqref{V_g} using the trained potential $V_g(z)$.}
    \label{bp_plot}
\end{figure}

\begin{figure}
    \centering
    \includegraphics[width=8cm]{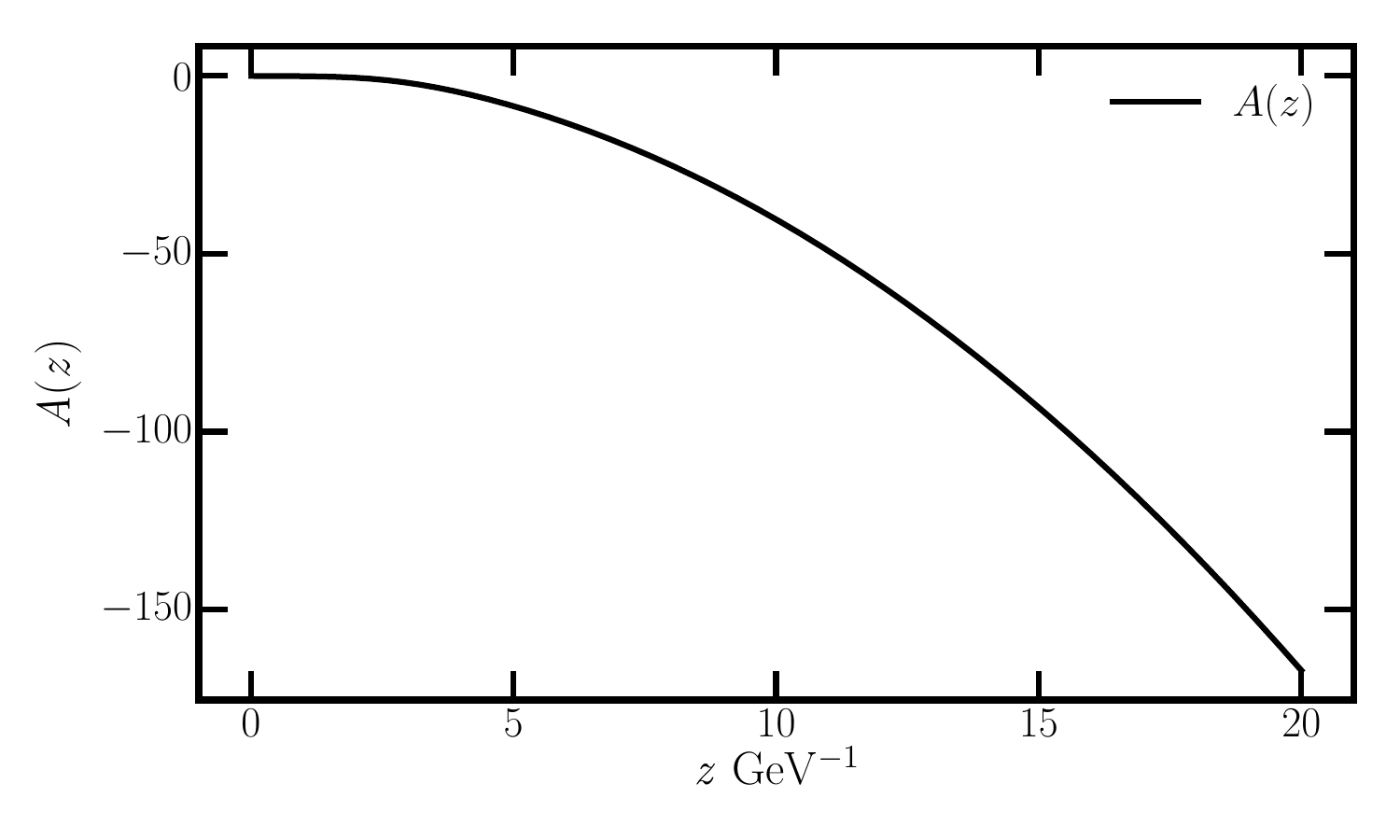}
    \caption{Final reconstructed warp factor $A(z)$ incorporating entropy and glueball constraints. }
    \label{A_plot}
\end{figure}

\begin{figure}[htbp]
    \centering
    \includegraphics[width=0.9\linewidth]{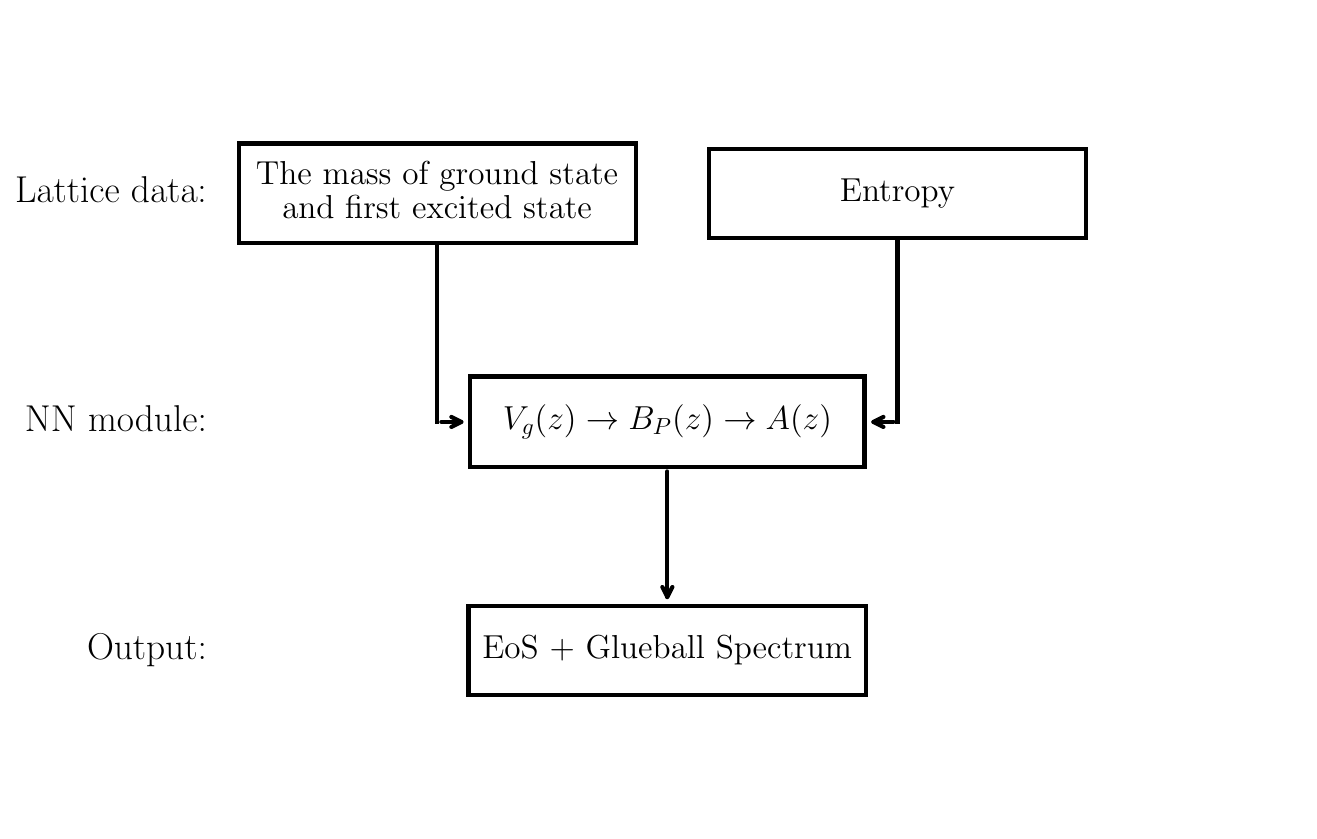}
    \caption{A sketch of holographic reconstruction workflow.}
    \label{flow_chart}
\end{figure}

\begin{figure}
    \centering
    \includegraphics[width=9cm]{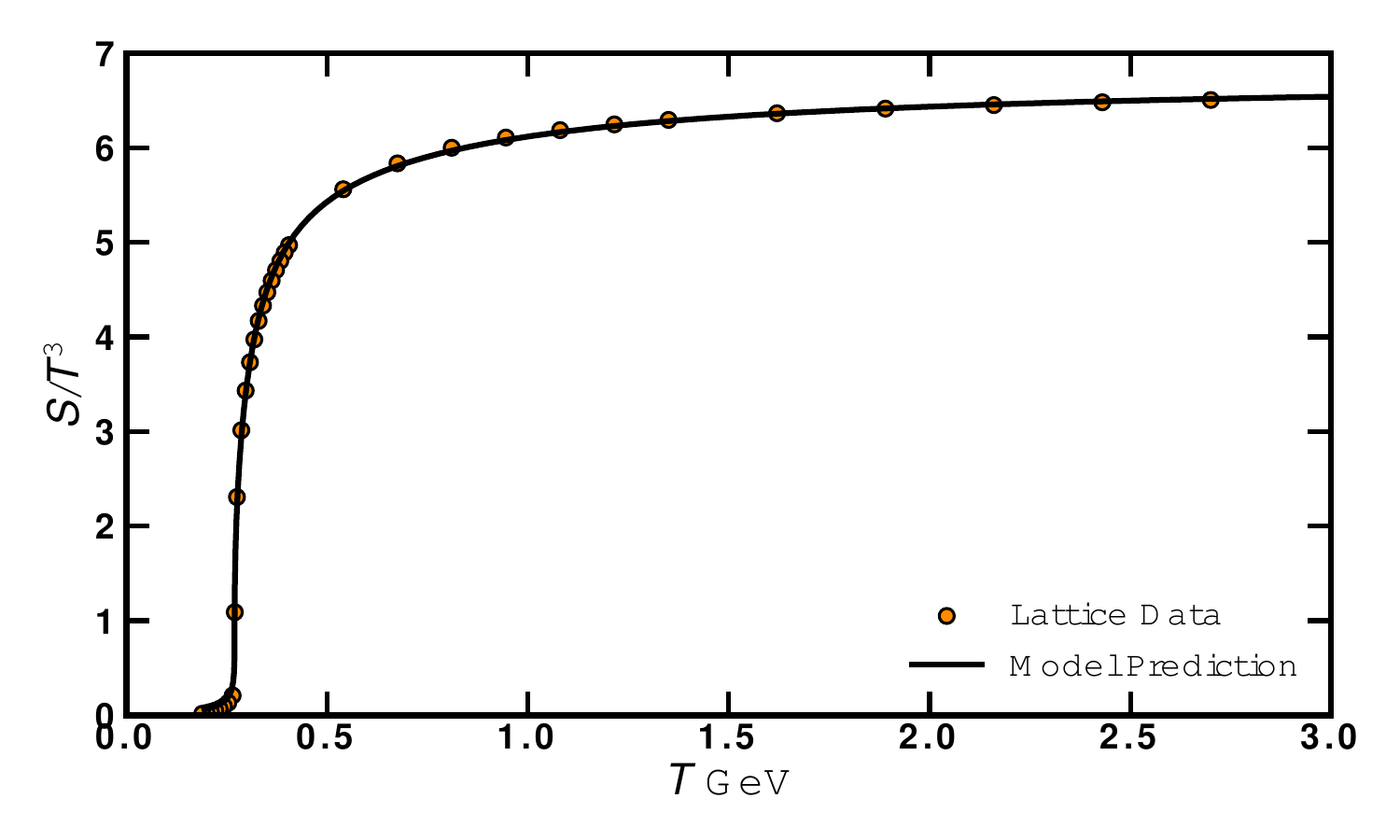}
    \caption{Entropy density $s/T^3$ versus temperature: model prediction using $A_1(z)$ compared with lattice QCD data \cite{Borsanyi:2012ve} ($G_5 = 1.16  \mathrm{GeV}^{-3}$).}
    \label{S_T3_plot}
\end{figure}

In a fully dynamical treatment, scalar glueball modes should arise from coupled metric–dilaton fluctuations in Refs. \cite{Gursoy:2007er, Ballon-Bayona:2017sxa}. In the present work, we adopt a simpler and widely used approximation in which glueballs are described by probe bulk fields propagating on a fixed background determined from thermodynamics. Under this assumption, the equations of motion decouple, which allows us to extract the glueball spectrum in a numerically efficient way and to directly connect it to the underlying data-driven dilaton potential. The 5D action for the scalar glueball field $\psi(x, z)$ in the string frame takes the same form as in the soft-wall model:
$$
S_{\psi} = \int \mathrm{d}^5 x \sqrt{g_s} \, e^{-\Phi} \left[ \frac{1}{2} \partial_M \psi \partial^M \psi + \frac{1}{2} M_5^2 \psi^2 \right],
$$
where $M_5$ is the 5-dimensional mass of the glueball. From the AdS/CFT dictionary, the scalar glueball has $M_5 = 0$. The metric in the string frame is characterized by the warp factor 
$$
A_s(z) = A(z) + \sqrt{\frac{1}{6}} \Phi(z) - \ln z.
$$
At vanishing temperature, the metric simplifies to
\begin{equation}
\mathrm{d} s^2 = e^{2 A_s(z)} \left( -\mathrm{d} t^2 + \mathrm{d} z^2 + \mathrm{d} \vec{x}^2 \right).
\end{equation}
The equation of motion for the scalar glueball field is
\begin{equation}
\partial_z^2 \psi_n + \left( 3A_s' - \Phi' \right) \partial_z \psi_n - m_n^2 \psi_n = 0,
\end{equation}
where we have defined $A_s' \equiv \partial_z A_s$ and $\Phi' \equiv \partial_z \Phi$. Applying the field transformation
\begin{equation}
\psi_n(z) = e^{-\frac{1}{2} \int \left( 3A_s'(\zeta) - \Phi'(\zeta) \right) \mathrm{d}\zeta} \widetilde{\psi}_n(z),
\end{equation}
we obtain the Schrödinger-like equation
\begin{equation}
-\widetilde{\psi}_n'' + V_g(z) \widetilde{\psi}_n = m_n^2 \widetilde{\psi}_n,
\end{equation}
with the effective potential
\begin{equation}\label{V_g}
V_g(z) = \frac{3A_s''(z) - \Phi''(z)}{2} + \frac{ \left( 3A_s'(z) - \Phi'(z) \right)^2 }{4}.
\end{equation}
To reconstruct the glueball potential $V_g(z)$ using lattice QCD data for the scalar glueball spectrum, we implement a neural network representation of $V_g(z)$ combining a physics-inspired base potential with a neural network correction term. The base potential consists of two learnable parameters $C_1$ and $C_2$ that create a potential of form $C_1/z^2 + C_2*z^2$. The $\frac{1}{z^2}$ term secures the analytic singularity structure near the boundary (dictated by operator dimensions and geometric contributions), while the $z^2$ term generates confining/soft-wall behavior in the infrared region to guarantee the confined nature of the spectrum \cite{Karch:2006pv}. The correction component is a 4-layer fully connected network with tanh activations taking a single input $z$ and passing it through sequential layers of 64 nodes each before outputting a single correction value. The network's final output sums the base potential and correction term to produce the complete potential function $V_g(z)$.  The reconstruction begins by constraining the potential with $m_0 = 1.475 \mathrm{GeV}$ (ground state) and $m_1 = 2.775 \mathrm{GeV}$ (first excited state) from lattice QCD \cite{Meyer:2004gx}. To numerically solve the Schr\"odinger equation, we employ a finite-difference discretization scheme on a uniform grid. For the wavefunction, we impose Dirichlet boundary conditions at both ends:$\widetilde{\psi}_n(z_{\min})=0,\quad \widetilde{\psi}_n(z_{\max})=0$. Through an iterative optimization process, we solve the Schrödinger equation numerically for the eigen masses $\{m_n^{\text{pred}}\}$, comparing them with experimental values. The neural network parameters are continuously refined by minimizing the loss function $\mathcal{L} = \sum_{n=0}^1 \left( m_n^{\text{pred}} - m_n^{\text{exp}} \right)^2$ until convergence is achieved ($\mathcal{L} \rightarrow 0$). We set an early-stop convergence criterion: the program will terminate and output the corresponding V-network once the loss function reaches $1 \times 10^{-7}$. The reconstructed potential $V_g(z)$ obtained through this spectral matching procedure is shown in Fig. \ref{potential_plot}. The qualitative behavior of the glueball potential $V_g(z)$ is consistent with predictions from other holographic QCD models.
We then define the composite derivative function $B_P(z) = 3A_s'(z) - \Phi'(z)$, allowing us to express the effective potential as
\begin{equation}\label{V_g}
V_g(z) = \frac{B_P'(z)}{2} + \frac{B_P(z)^2}{4}.
\end{equation}
This $B_P(z)$ function is represented by a new neural network. It uses a 1-128-128-1 fully connected structure with SiLU activation functions between layers, chosen for their smooth gradients and nonlinear properties. The network first processes the input $z$ through these hidden layers to produce a raw unbounded output. This output then undergoes an absolute value transformation with a small epsilon offset (1e-6) to ensure strict positivity before being negated in the final step, guaranteeing all outputs will be negative while maintaining differentiability. Using the previously reconstructed $V_g(z)$ (shown in Fig. \ref{potential_plot}) as the target solution, we solve the differential equation \eqref{V_g} to obtain $B_P(z)$. The resulting function is displayed in Fig. \ref{bp_plot}.
Within this framework, $\Phi(z)$ and $A(z)$ are fundamentally related through the constraint \cite{Chen:2024ckb,Chen:2024mmd,Zhu:2025gxo}:
\begin{equation}
\Phi'(z) = \sqrt{6 \left[ (A'(z))^2 - A''(z) - \frac{2A'(z)}{z} \right] }.
\end{equation}
To solve this system, we represent $A(z)$ through a new neural network which is a 4-layer fully-connected neural network (1-64-128-64-1) with Sigmoid activation in all layers. The output is specially processed: first constrained to [0,1] by the final Sigmoid, then scaled to negative values by multiplying with -10, and finally multiplied with input $z$. This design enforces a negative output and $A(0) = 0$. To further constrain the model, we conduct a dedicated training of the neural network for $A(z)$ using lattice QCD entropy density data ($s/T^3$), treating $G_5$ as a free parameter. This optimization yields the Newton constant $G_5 = 1.16 \,\mathrm{GeV}^{-3}$ and produces a solution denoted $A_1(z)$ (previously referred to as $A(z)$) that accurately reproduces the QCD entropy above $T_c$.
We then reconcile this entropy-constrained $A_1(z)$ with the glueball potential-derived $B_P(z)$ in a unified reconstruction of final $A(z)$. The network $A(z)$ is a four-layer fully connected neural network with input and output dimensions of 1 and two hidden layers of size 64. All weights are forced to be positive during initialization, and tanh activation is applied to the first three layers to ensure monotonicity. The output layer applies a linear transformation followed by a negative exponential, guaranteeing strictly negative and monotonically decreasing outputs. The final optimization of \( A(z) \) incorporates three complementary constraints: matching \( B_P(z) \) in the IR region \( z \in [4, 20]\,\mathrm{GeV}^{-1} \), matching \( A_1(z) \) in the UV region \( z \in [0.1, 2]\,\mathrm{GeV}^{-1} \), and leaving a strategically placed buffer zone \( z \in [2, 4]\,\mathrm{GeV}^{-1} \) unconstrained to enable smooth interpolation between the UV and IR regimes. The resulting warp factor $A(z)$ is shown in Fig. \ref{A_plot}. 
To conclude this part, the holographic reconstruction workflow progresses through sequential stages of training as summarized in Fig. \ref{flow_chart}. The procedure initiates by transforming glueball spectra data into the reconstructed metric potential $V_g(z)$ via Schrödinger inversion. Building upon this foundation, the composite derivative function $B_P(z) = 3A_s'(z) - \Phi'(z)$ is then extracted through potential decomposition by solving Eq. \eqref{V_g}. The workflow culminates in integrating EoS constraints with the derived $B_P(z)$ to reconstruct the fundamental warp factor $A(z)$. With the neural-network function $A(z)$, constrained by the equation of state and the glueball spectrum, we computed the entropy density and compared it with lattice results, which show good agreement. At low temperatures, $s/T^{3}$ approaches zero, while at high temperatures it tends to a constant, as shown in Fig.~\ref{S_T3_plot}.

\section{ EoS and spectrum}
\begin{figure}
    \centering
    \includegraphics[width=8cm]{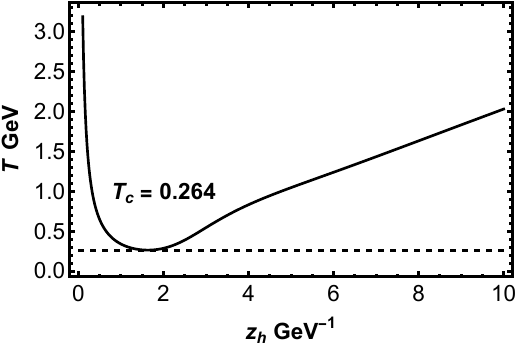}
    \caption{The temperature $T$ as a function of $z_h$. $T_c$ is the phase transition temperature.}
    \label{temperature}
\end{figure}

\begin{figure}
    \centering
    \includegraphics[width=8cm]{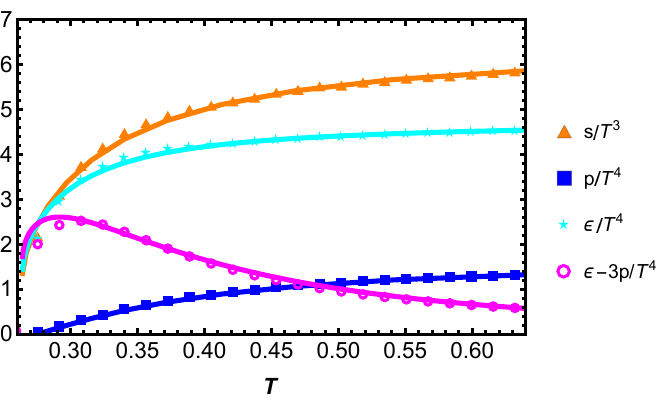}
    \caption{Thermodynamics of pure Yang-Mills theory: Holographic model predictions (solid lines) versus lattice QCD data (symbols) above $T_c = 0.264$ GeV \cite{Borsanyi:2012ve}.}
    \label{pureeos}
\end{figure}

Having reconstructed the Einstein-Dilaton model via neural network optimization, we now validate our framework and present predictive capabilities. First, we compute the scalar glueball mass spectra using the reconstructed gravitational dual. As shown in Table \ref{table1}, our predictions for the $0^{++}$ states exhibit strong agreement with lattice QCD results across multiple calculations. 
\begin{table}
\centering
\begin{tabular}{c|cccc}
State & Lat1 & Lat2 & Lat3 & Our model \\ \hline
$0^{++}$  & 1475(30)(65) & 1653(26) & 1730(50)(80) & 1791 \\
$0^{++*}$  & 2755(70)(120) & 2842(40) & 2670(180)(130) & 2751 \\
$0^{++**}$  & 3370(100)(150) & -- & -- & 3500 \\
$0^{++***}$  & 3990(210)(180) & -- & -- & 4122 \\ 
\end{tabular}
\caption{Scalar glueball spectrum ($J^{PC} = 0^{++}$) comparisons. Masses in MeV. Lattice data sources: Lat1~\cite{Meyer:2004gx}, Lat2~\cite{Athenodorou:2020ani}, Lat3~\cite{Morningstar:1999rf}.}
\label{table1}
\end{table}
In Fig. \ref{temperature}, we plot the temperature as a function of the horizon position $z_h$. The confinement-deconfinement transition temperature is found to be $T_c = 0.264\, \rm GeV$.
We further compute the EoS for pure Yang-Mills theory within our holographic framework. As shown in Fig. \ref{pureeos}, the reconstructed model accurately reproduces key thermodynamic quantities. The pressure ($p/T^4$), energy density ($\epsilon/T^4$), and trace anomaly ($(\epsilon-3p)/T^4$) all exhibit excellent agreement with lattice QCD results above $T_c$.

\section{Conclusion and outlook} We have developed a neural network framework to solve the inverse problem of bottom-up holography: starting from lattice inputs for the ground and first-excited $0^{++}$ glueball masses plus the entropy density, we reconstructed a self-consistent Einstein-dilation background that simultaneously captures confinement thermodynamics and spectroscopy. The resulting Neural-Network Holographic model reproduces lattice results for the pressure, energy density and trace anomaly above the deconfinement temperautre with high fidelity and predicts the $0^{++**}$ and $0^{++***}$ masses in consistent with available lattice calculations. This unified, data-driven construction establishes a quantitative bridge between gauge-theory observables and their gravitational duals. Looking ahead, this framework shows promise for extension across several research directions: comprehensive spectroscopic studies of higher-spin glueballs ($2^{++}$, $1^{+-}$) and their excitations; investigations of finite-density physics including chemical potential effects and critical endpoints; computations of real-time dynamics such as transport coefficients; A fully dynamical approach to the equation of state and the glueball spectrum can be investigated in future work. These developments may contribute to improved computational approaches for non-perturbative phenomena while helping to deepen understanding of connections between emergent gravity and quantum chromodynamics, potentially enabling new perspectives on strongly-coupled matter systems. 

\section*{Acknowledgments}
 This work is supported in part by the National Natural Science Foundation of China (NSFC) Grant Nos: 12405154 and 12305136, the start-up funding of Hangzhou Normal University under Grant No.\ 4245C50223204075, the European Union -- Next Generation EU through the research grant number P2022Z4P4B ``SOPHYA - Sustainable Optimised PHYsics Algorithms: fundamental physics to build an advanced society'' under the program PRIN 2022 PNRR of the Italian Ministero dell'Universit\`a e Ricerca (MUR), the CUHK-Shenzhen university development fund under grant No.\ UDF01003041 and UDF03003041, and Shenzhen Peacock fund under No.\ 2023TC0179.
 
\section*{References}

\bibliography{ref}
\end{document}